\documentclass{article}
\usepackage{graphicx}
\usepackage{rotating}
\usepackage[table]{xcolor}
\definecolor{lightgray}{gray}{0.9}

\begin{document}

\title{Further Insights into the Interareal Connectivity of a Cortical Network}

\author{Luciano~Dyballa\\
Valmir~C.~Barbosa\thanks{Corresponding author (valmir@cos.ufrj.br).}\\
\\
Programa de Engenharia de Sistemas e Computa\c c\~ao, COPPE\\
Universidade Federal do Rio de Janeiro\\
Caixa Postal 68511\\
21941-972 Rio de Janeiro - RJ, Brazil}

\date{}

\maketitle

\begin{abstract}
Over the past years, network science has proven invaluable as a means to better understand many of the processes taking place in the brain. Recently, interareal connectivity data of the macaque cortex was made available with great richness of detail. We explore new aspects of this dataset, such as a correlation between connection weights and cortical hierarchy. We also look at the link-community structure that emerges from the data to uncover the major communication pathways in the network, and moreover investigate its reciprocal connections, showing that they share similar properties.

\end{abstract}

\newpage
\section{Introduction}\label{intro}
Recent years have seen an increasing use of tools from network science as a means to make sense of the staggering complexity of the brain and to uncover some of the mechanisms governing its organization \cite{bullmore09,stam12}. The view of the central nervous system as being formed by complex networks at a number of different scales has been largely adopted in neuroscience, spurring the interest of neuroscientists in graph-theoretic methods to characterize the structural and functional connectivity patterns between regions of the brain \cite{sporns10}.

Approaches based on this view have succeeded in revealing many different aspects of brain organization and in providing important clues of the functional specialization of cortical areas \cite{bressler10}. Several laboratories have thus been stimulated to undertake the effort of mapping the large-scale networks in the central nervous systems of humans \cite{hagmann08}, monkeys \cite{fve91}, cats \cite{scannell95}, and mice \cite{oh14}, providing for ever more accurate and complete databases.

Interareal structural networks represent the interactions between different regions in the brain. Each node in the network corresponds to a cytoarchitecturally defined area and each link (or edge) represents a physical connection promoted by the axonal pathways between two areas. Connectivity data of this kind allow us to investigate properties of the high-level processes taking place in the cortex, such as communication efficiency  \cite{ercsey13}, integration  of information \cite{zamora10,vandenheuvel13}, modular organization \cite{meunier10}, robustness against lesion \cite{kaiser07}, and the effect of diseases in connectivity \cite{bassett08}.

A recent study \cite{markov12} has used quantitative anatomical tract tracing to map the interareal connectivity of the macaque monkey cerebral cortex with unprecedented richness of detail. In contrast with other widely studied datasets \cite{scannell99,modha10}, the new data include not only the direction, but also the number of neurons involved in each connection, as well as the laminar origin distribution for many of them \cite{markov14hierarchy}. Links with this kind of additional information may enable us to take into account important differences among connections and thus better understand the system as a whole \cite{newman04}. This dataset is also particularly remarkable because of its high degree of reliability, since all the tracing experiments were conducted by the same group and therefore were subject to the same criteria and statistical validation \cite{markov12}.

The present work attempts to further explore this still largely uncharted territory of the interareal network of the macaque cortex. We investigate properties and correlations that have not been looked into before by making use of the particularities of the new data, such as the number of neurons involved in each connection, and present new perspectives from which to uncover the regularities in the complex networks of the brain. In particular, we present a correlation between weight distribution and hierarchical level per cortical area, an analysis of the link-community structure of the network, as well as correlations between the number of neurons in reciprocal connections.

We begin in Section~\ref{dataset} by describing the dataset and defining some of the properties utilized to characterize the links in the network. In Section~\ref{hierarchy}, we show that there is a correlation between each area's incoming weight distribution and its position in the cortical hierarchy. Section~\ref{linkcomm} presents the organization of the network into link communities, revealing the main paths of communication between regions in the macaque cortex. We proceed to investigate correlations between reciprocal links, which constitute the majority of connections in the network, in Section~\ref{reciprocals}. A discussion of our results is found in Section~\ref{discussion}, followed by conclusions in Section~\ref{conclusion}.

\section{Dataset characteristics and definitions}\label{dataset}

The authors of \cite{markov12} used anatomical tract tracing, employing retrograde tracers, to map the interareal connections in the macaque cortex. In this kind of procedure, the tracer is injected in a given target area and subsequently diffuses along the axons that terminate in that area, traveling back to the neurons from which the axonal projections originate. According to a parceling scheme, the locations of these neurons are matched to known cytoarchitecturally defined areas. The areas labeled in this way are then included in the data as in-neighbors of the injected area. This particular experiment consisted of repeated injections in 29 cortical areas spanning the four lobes in the left hemisphere of the macaque cortex, out of a total of 91 areas \cite{markov12}.

It is important to note that, even though we only have 29 injected areas, the data include the connections arriving at them from all 91 areas. As a consequence, we have virtually all connections that exist among the 29 injected areas, making for a $29\times29$ adjacency matrix. The resulting network contains 536 directed edges. We can also consider the $91\times29$ incomplete adjacency matrix, which includes all connections detected in the experiment.

We proceed by presenting some of the terminology used to characterize each projection. The complete dataset and more information can be found at http://www.core-nets.org.

The number of neurons (NN) of a given connection from area B to area A corresponds to the number of neurons labeled in B after tracer injection in A. The NN value used for this connection is the geometric mean of the values for all subjects.

The quantity used in \cite{markov12} as the weight of an edge from area B to area A is the fraction of labeled neurons (FLN) in area B relative to all neurons labeled upon tracer injection in area A. The FLN is therefore a normalized version of the NN and is useful because it helps to assess the relative contribution of each connection to the area receiving it, irrespective of the area's volume or cell density.

The dataset also includes approximations to the axonal distance between areas. Throughout the paper we refer to the resulting values as connection lengths.

\section{Cortical hierarchy}\label{hierarchy}

\subsection{Weight distribution per area}

Because of the way the FLN weights are defined, the incoming weights for any of the 29 injected areas sum up to 1 when we consider incoming connections from all 91 cortical areas. This is expected to remain the same in a future dataset based on injections in all 91 areas, so already at this point it is worth investigating how these weights are distributed.

Table~\ref{table-flndistr} shows the FLN distribution for the ten connections with highest FLN in four different areas: V1, 2, 7A, and 10. One aspect that is common to all of them is that they all have few connections with relatively large FLN and a large number of connections with very small FLN (say, smaller than 1\%). Interestingly, this same pattern is seen for all the 29 injected areas, which means that, for instance, there is no single area having its total NN evenly distributed among its in-neighbors.

\begin{table}[t]
\centering
\caption{Ten highest FLN values of the connections incoming to areas V1, 2, 7A, and 10. The bottom row contains the mean FLN value over all incoming connections.}
\label{table-flndistr}
\rowcolors{1}{white}{lightgray}

\begin{tabular}{ c c c c}

\hline 

\textbf{V1} & \textbf{2} & \textbf{7A} & \textbf{10} \\
\hline
0.7313 &  0.3800 &  0.1640 &  0.2042 \\
0.1196 &  0.2400 &  0.1290 &  0.1796 \\
0.0581 &  0.1420 &  0.1260 &  0.1086 \\
0.0235 &  0.1110 &  0.0921 &  0.0815 \\
0.0073 &  0.0424 &  0.0783 &  0.0638 \\
0.0066 &  0.0300 &  0.0774 &  0.0337 \\
0.0055 &  0.0104 &  0.0733 &  0.0332 \\
0.0043 &  0.0093 &  0.0413 &  0.0277 \\
0.0035 &  0.0078 &  0.0364 &  0.0242 \\
0.0030 &  0.0076 &  0.0211 &  0.0238 \\
$\cdots$ & $\cdots$ & $\cdots$ & $\cdots$ \\
\hline
Mean: 0.0963 & Mean: 0.0980 & Mean: 0.0839 & Mean: 0.0780 \\
\hline
\end{tabular}

\end{table}

But despite this similarity, there are some marked differences between the areas' weight distributions. Some areas, like V1, have one or two incoming connections with very high FLN, followed by several ones with much smaller FLN, while others, such as 7A, seem to have a much less pronounced variation of FLN among its in-neighbors. The fact that V1 is an area lower than 7A in terms of a hierarchy of information processing in the cortex seems to indicate that each area's weight distribution might tell us something about that area's hierarchical position.

\subsection{Hierarchical distance}

To investigate whether the FLN distribution of an area can tell us something about that area's position in the cortical hierarchy, we define the hierarchical distance, HD, of a given area A to be the smallest directed distance to it from one of the sensory input areas in the cortex---namely V1 (primary visual cortex); 1, 2, and 3 (primary somatosensory cortex); Gu (primary gustative cortex); ENTO and PIRI (primary olfactory cortex); and Core (primary auditory cortex). Hierarchical distances, therefore, are relative to the $91\times29$ adjacency matrix.

Each edge is assigned a length equal to the inverse of its FLN weight (which means that the larger an edge's FLN, the smallest the length between the two nodes it connects). Hence, the directed distance from B to A is the total length of the directed path from B to A whose total length is minimum. The value of HD for all sensory input areas is therefore 0.

For example, the value of HD for V2 is $1.31$, since it has a single-edge path of length $1/0.76 = 1.31$ from V1 and no shorter path from any of the other sensory input areas. The greater an area's hierarchical distance, the higher it is in the cortical hierarchy.

Other metrics for hierarchical distance have been proposed \cite{fve91,hilgetag10}, taking into consideration the laminar distributions at the origin and termination of each connection, which are not available in the dataset at hand. Nevertheless, Table~\ref{table-hd} shows that our method ranks the 29 injected areas in an order that is roughly in accordance with other rankings reported for the macaque cortex \cite{barone00,reid09}, indicating that we have a reasonable, though approximate, measure.

\begin{table}[t]
\centering
\rowcolors{1}{white}{lightgray}
\caption{Hierarchical distance (HD) for the 29 injected areas.}
\label{table-hd}

\begin{tabular}{ c c}

\hline 

\textbf{Area} & \textbf{HD} \\
\hline
9/46v & 60.48 \\
7m & 45.68 \\
10 & 45.41 \\
46d & 39.84 \\
7A & 39.57 \\
9/46d & 38.19 \\
8l & 36.09 \\
24c & 34.50 \\
8m & 33.68 \\
7B & 32.25 \\
8B & 31.05 \\
F7 & 24.42 \\
STPc & 17.74 \\
STPr & 17.33 \\
STPi & 15.35 \\
TEpd & 15.04 \\
F2 & 14.13 \\
PBr & 12.18 \\
5 & 10.97 \\
F1 & 8.00 \\
TEO & 7.79 \\
DP & 6.84 \\
MT & 6.51 \\
ProM & 6.21 \\
F5 & 4.93 \\
V4 & 3.88 \\
V2 & 1.31 \\
V1 & 0.00 \\
2 & 0.00 \\
\hline

\end{tabular}

\end{table}

\subsection{Hierarchical distance \textit{vs}.\ mean FLN}

\begin{figure}[t]
\centering
\scalebox{0.80}{\includegraphics{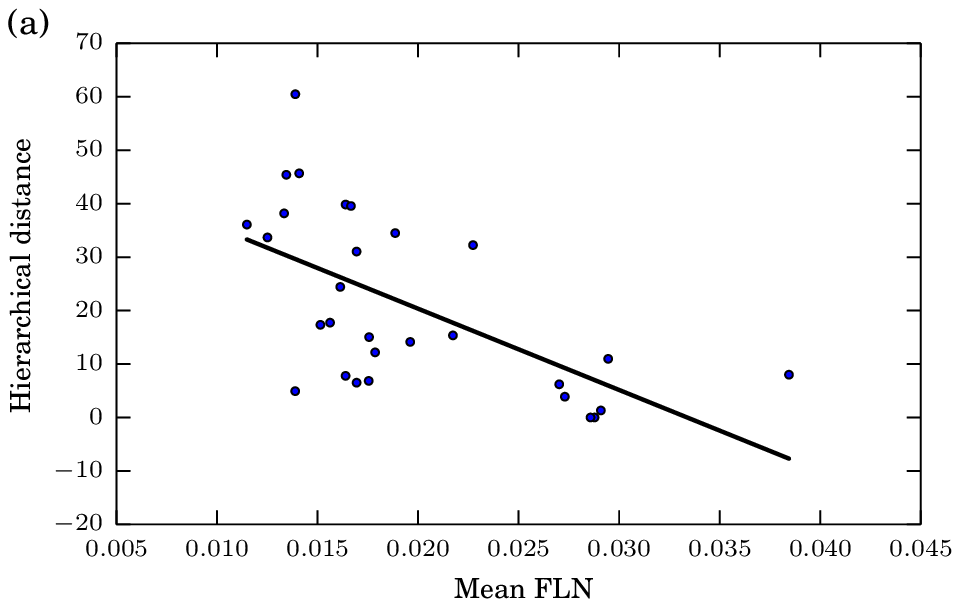}}
\scalebox{0.80}{\includegraphics{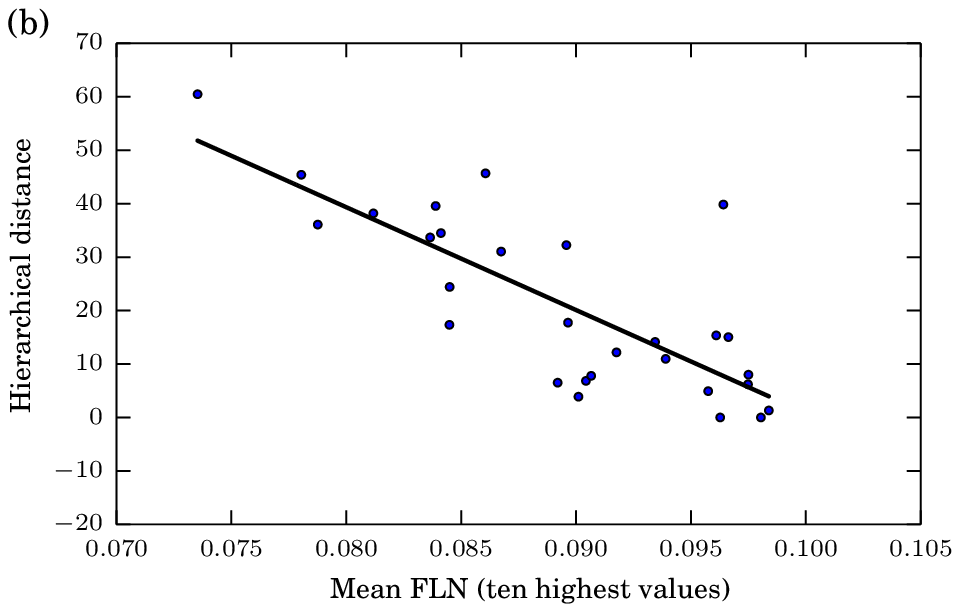}}
\caption{Hierarchical distance (HD) \textit{vs.}\ mean FLN value of all incoming connections (a), and HD \textit{vs.}\ mean FLN value when considering only the ten highest FLN values (b). Each point corresponds to one of the 29 injected areas.}
\label{fig-hierarch}
\end{figure}

When we plot each area's hierarchical distance versus its mean FLN value (Figure~\ref{fig-hierarch}(a)), we find that the two quantities are strongly negatively correlated (Pearson correlation coefficient $r = -0.61$). This suggests that, by looking at the way the connection strengths are distributed among an area's in-neighbors, it may be possible to tell whether it processes high- or low-level information. Interestingly, when we take the mean of the ten connections with highest FLN for each area (Figure~\ref{fig-hierarch}(b)), the linearity is even greater ($r = -0.79$).

\section{Link communities}\label{linkcomm}

The community structure, or modular organization, of the mammalian cortex has been largely investigated \cite{meunier10}. It is supposed to promote functional segregation by means of a high degree of interaction between areas sharing similar functional roles \cite{sporns13segregation}, forming modules or communities. These, in turn, facilitate global integration through the communication between hubs pertaining to different communities \cite{vandenheuvel13,vandenheuvel13hubs}.

Most studies of brain networks have used traditional node-community detection techniques, which partition the network into the modules or communities that yield high intramodular and low intermodular edge densities. The major drawback of such an approach is that the resulting communities cannot overlap, i.e., each node belongs to a single community. Previous results have found node communities highly related to the physical positions of their constituting areas \cite{goulas14}, with each community corresponding roughly to the cortical region where its areas are located. These communities do not reveal much more than what one would naturally expect, however, since nearby areas usually share similar functions. Also, since shorter connections tend to have higher FLN values \cite{markov12}, it is likely that methods for community detection based on modularity maximization will group closely positioned areas in the macaque cortex into the same community.

Actually, we would like a community structure to provide us with insight about functional similarity, irrespective of the strengths of connections. The weights should not be ignored, however, since we cannot overlook the fact that a weak long-distance connection is highly different from a strong short one. Furthermore, given the highly integrative character of the cortex, is seems natural to think that each module is not isolated, and should probably have one or more nodes responsible for the exchange of information with other modules with distinct functional roles. We find that the detection of link communities \cite{ahn10} is a natural way to incorporate this perspective into our analysis, and more: by grouping links instead of nodes, we expect to capture more meaningful communities that not only will tell us which nodes are more related to which others, but also the nature of their relations, as indicated by the directions of the links in each community.

To accomplish this, we have used the method described in \cite{ahn10}, which involves calculating a similarity measure based on neighborhood overlap for all pairs of links having a node in common. A hierarchical clustering algorithm is subsequently applied to build a link dendrogram, whose section with maximum partition density is the one with heuristically better communities. In this study, we have made a small adaptation to the similarity formula to better incorporate weights and directions and to allow for reciprocal connections (see Appendix~\ref{methods} for details).

\begin{figure}[t]
\centering
\scalebox{0.80}{\includegraphics{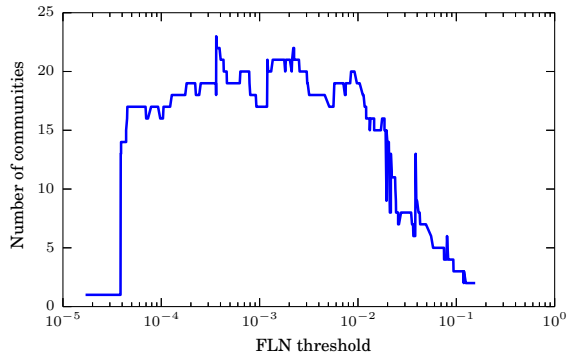}}
\caption{Number of communities found for different FLN thresholds. A maximum of 23 communities is found for a threshold of 0.000362.}
\label{fig-nofcomms}
\end{figure}

Applying single-linkage hierarchical clustering when using all links implied by the $29\times29$ adjacency matrix, however, results in a poor-quality community structure, with a single module that includes all nodes. The results change dramatically, though, if we filter out the weakest links. We looked for the FLN threshold that yields the maximum number of link communities, and found it to be 0.000362 (Figure~\ref{fig-nofcomms}). This means that, by using only links with $\textrm{FLN} \geq 0.000362$ (in other words, discarding all links $\textrm{B}\rightarrow\textrm{A}$ whose NN is less than 0.0362\% of the total NN projecting to A), we uncover a partition of the links into 23 distinct communities. (See Section~\ref{discussion} for a more thorough discussion concerning the filtering of links.)

\begin{figure}[p]
\centering
\scalebox{0.60}{\includegraphics{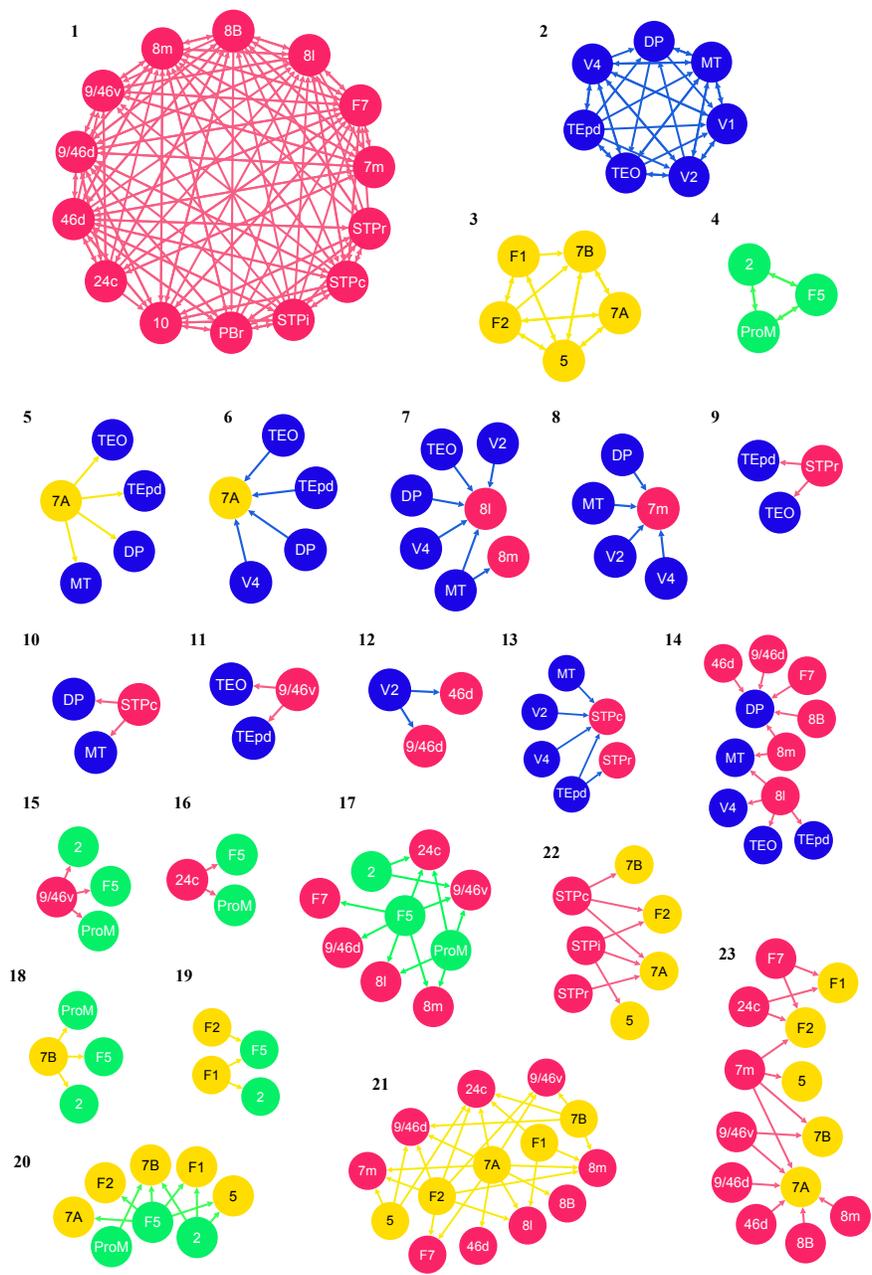}}
\caption{Link communities found using the method described in Appendix~\ref{methods}. Colors used in the nodes of communities 5--23 refer to the colors used in the first four communities. Links are colored as the node where they originate.}
\label{fig-comms}
\end{figure}

The communities found are shown in Figure~\ref{fig-comms}. We also give, in Table~\ref{table-pcts}, a list of each area's percentage of contribution to each community, defined as the fraction of links incident to an area A in a given community relative to all the links that are incident to A.

\begin{sidewaystable}
\scriptsize
\centering
\caption{Percentage of participation of each area in each community of Figure~\ref{fig-comms}.}
\rowcolors{1}{white}{lightgray}
\begin{tabular}{l c c c c c c c c c c c c c c c c c c c c c c c}

\hline 
& \multicolumn{23}{c}{\textbf{Community}} \\
\cline{2-24}
\textbf{Area}  & \textbf{1}  & \textbf{2}  & \textbf{3}  & \textbf{4}  & \textbf{5}  & \textbf{6}  & \textbf{7}  & \textbf{8}  & \textbf{9}  & \textbf{10}  & \textbf{11}  & \textbf{12}  & \textbf{13}  & \textbf{14}  & \textbf{15}  & \textbf{16}  & \textbf{17}  & \textbf{18}  & \textbf{19}  & \textbf{20}  & \textbf{21}  & \textbf{22}  & \textbf{23}  \\
\hline
\textbf{10}  & 100.0   & -   & -   & -   & -   & -   & -   & -   & -   & -   & -   & -   & -   & -   & -   & -   & -   & -   & -   & -   & -   & -   & -  \\
\textbf{2}   & -   & -   & -  & 33.4   & -   & -   & -   & -   & -   & -   & -   & -   & -   & -  & 8.3   & -  & 16.7  & 8.3  & 8.3  & 25.0   & -   & -   & -  \\
\textbf{24c}  & 55.6   & -   & -   & -   & -   & -   & -   & -   & -   & -   & -   & -   & -   & -   & -  & 7.4  & 11.1   & -   & -   & -  & 18.5   & -  & 7.4  \\
\textbf{46d}  & 85.6   & -   & -   & -   & -   & -   & -   & -   & -   & -   & -  & 3.6   & -  & 3.6   & -   & -   & -   & -   & -   & -  & 3.6   & -  & 3.6  \\
\textbf{5}   & -   & -  & 53.3   & -   & -   & -   & -   & -   & -   & -   & -   & -   & -   & -   & -   & -   & -   & -   & -  & 13.3  & 20.0  & 6.7  & 6.7  \\
\textbf{7A}   & -   & -  & 18.2   & -  & 12.1  & 12.1   & -   & -   & -   & -   & -   & -   & -   & -   & -   & -   & -   & -   & -  & 3.0  & 27.3  & 9.1  & 18.2  \\
\textbf{7B}   & -   & -  & 31.5   & -   & -   & -   & -   & -   & -   & -   & -   & -   & -   & -   & -   & -   & -  & 15.8   & -  & 15.8  & 21.1  & 5.3  & 10.5  \\
\textbf{7m}  & 57.7   & -   & -   & -   & -   & -   & -  & 15.4   & -   & -   & -   & -   & -   & -   & -   & -   & -   & -   & -   & -  & 11.5   & -  & 15.4  \\
\textbf{8B}  & 88.0   & -   & -   & -   & -   & -   & -   & -   & -   & -   & -   & -   & -  & 4.0   & -   & -   & -   & -   & -   & -  & 4.0   & -  & 4.0  \\
\textbf{8l}  & 57.5   & -   & -   & -   & -   & -  & 15.2   & -   & -   & -   & -   & -   & -  & 12.1   & -   & -  & 6.1   & -   & -   & -  & 9.1   & -   & -  \\
\textbf{8m}  & 69.7   & -   & -   & -   & -   & -  & 3.0   & -   & -   & -   & -   & -   & -  & 6.1   & -   & -  & 6.1   & -   & -   & -  & 12.1   & -  & 3.0  \\
\textbf{9/46d}  & 74.3   & -   & -   & -   & -   & -   & -   & -   & -   & -   & -  & 3.2   & -  & 3.2   & -   & -  & 3.2   & -   & -   & -  & 12.9   & -  & 3.2  \\
\textbf{9/46v}  & 55.3   & -   & -   & -   & -   & -   & -   & -   & -   & -  & 6.9   & -   & -   & -  & 10.3   & -  & 10.3   & -   & -   & -  & 10.3   & -  & 6.9  \\
\textbf{DP}   & -  & 41.1   & -   & -  & 5.9  & 5.9  & 5.9  & 5.9   & -  & 5.9   & -   & -   & -  & 29.4   & -   & -   & -   & -   & -   & -   & -   & -   & -  \\
\textbf{F1}   & -   & -  & 35.7   & -   & -   & -   & -   & -   & -   & -   & -   & -   & -   & -   & -   & -   & -   & -  & 14.3  & 14.3  & 21.4   & -  & 14.3  \\
\textbf{F2}   & -   & -  & 33.3   & -   & -   & -   & -   & -   & -   & -   & -   & -   & -   & -   & -   & -   & -   & -  & 4.8  & 4.8  & 33.3  & 9.5  & 14.3  \\
\textbf{F5}   & -   & -   & -  & 20.0   & -   & -   & -   & -   & -   & -   & -   & -   & -   & -  & 5.0  & 5.0  & 30.0  & 5.0  & 10.0  & 25.0   & -   & -   & -  \\
\textbf{F7}  & 79.4   & -   & -   & -   & -   & -   & -   & -   & -   & -   & -   & -   & -  & 3.4   & -   & -  & 3.4   & -   & -   & -  & 6.9   & -  & 6.9  \\
\textbf{MT}   & -  & 57.8   & -   & -  & 5.3   & -  & 10.5  & 5.3   & -  & 5.3   & -   & -  & 5.3  & 10.5   & -   & -   & -   & -   & -   & -   & -   & -   & -  \\
\textbf{PBr}  & 100.0   & -   & -   & -   & -   & -   & -   & -   & -   & -   & -   & -   & -   & -   & -   & -   & -   & -   & -   & -   & -   & -   & -  \\
\textbf{ProM}   & -   & -   & -  & 33.4   & -   & -   & -   & -   & -   & -   & -   & -   & -   & -  & 8.3  & 8.3  & 33.4  & 8.3   & -  & 8.3   & -   & -   & -  \\
\textbf{STPc}  & 69.0   & -   & -   & -   & -   & -   & -   & -   & -  & 6.9   & -   & -  & 13.8   & -   & -   & -   & -   & -   & -   & -   & -  & 10.3   & -  \\
\textbf{STPi}  & 84.2   & -   & -   & -   & -   & -   & -   & -   & -   & -   & -   & -   & -   & -   & -   & -   & -   & -   & -   & -   & -  & 15.8   & -  \\
\textbf{STPr}  & 78.9   & -   & -   & -   & -   & -   & -   & -  & 10.5   & -   & -   & -  & 5.3   & -   & -   & -   & -   & -   & -   & -   & -  & 5.3   & -  \\
\textbf{TEO}   & -  & 62.8   & -   & -  & 6.2  & 6.2  & 6.2   & -  & 6.2   & -  & 6.2   & -   & -  & 6.2   & -   & -   & -   & -   & -   & -   & -   & -   & -  \\
\textbf{TEpd}   & -  & 53.2   & -   & -  & 6.7  & 6.7   & -   & -  & 6.7   & -  & 6.7   & -  & 13.3  & 6.7   & -   & -   & -   & -   & -   & -   & -   & -   & -  \\
\textbf{V1}   & -  & 100.0   & -   & -   & -   & -   & -   & -   & -   & -   & -   & -   & -   & -   & -   & -   & -   & -   & -   & -   & -   & -   & -  \\
\textbf{V2}   & -  & 66.6   & -   & -   & -   & -  & 6.7  & 6.7   & -   & -   & -  & 13.3  & 6.7   & -   & -   & -   & -   & -   & -   & -   & -   & -   & -  \\
\textbf{V4}   & -  & 69.0   & -   & -   & -  & 6.2  & 6.2  & 6.2   & -   & -   & -   & -  & 6.2  & 6.2   & -   & -   & -   & -   & -   & -   & -   & -   & -  \\
\hline
\end{tabular}

\label{table-pcts}
\end{sidewaystable}

%\begin{table}[p]
%\centering
%\rotcaption{Percentage of participation of each area in each community of Figure~\ref{fig-comms}.}
%\rowcolors{1}{white}{lightgray}
%\begin{sideways}
%\input{table-pcts}
%\end{sideways}
%\label{table-pcts}
%\end{table}

The first four communities stand out in that they are highly clustered, each containing almost all possible links between their nodes. Together, they cover all 29 areas in the network without overlap, exactly like a partition of the nodes. Interestingly, these four communities resemble the ones previously reported using node-community detection \cite{goulas14}---in our results, we have one community too few and area 7m belongs to a different community. This fact alone indicates that the results obtained for link communities provide us with richer information, since they yield approximately the same modules obtained for node communities and even more. Furthermore, these four communities provide us with an easier way to look at the remaining 19, since the latter represent groups of links responsible for the interactions taking place among the former.

For instance, according to the numbering adopted in Figure~\ref{fig-comms}, community 7 tells us that area 8l is a major integrator of the information coming from the visual areas in community 2. Communities 5 and 6 indicate that area 7A functions as a mediator between the visual areas from community 2 and the parietal areas in community 3, since it acts both as an integrator of visual information (much like area 8l in community 7, but receiving projections from a slightly different set of areas) and as a disseminator of information influencing the visual areas, probably in the form of a feedback response since the areas in community 3 are higher in the cortical hierarchy (see Table~\ref{table-hd}). The same kind of analysis can be carried out for all the other communities.

Using the first four communities (1--4) as reference, we can summarize the major flows of information in the macaque cortex in the following way: communities 1 and 3 both send and receive signals to and from communities 2 and 4. The former two also exchange signals between themselves, in both directions. However, there is no community representing interactions between communities 2 and 4, which contribute to a view of these two subsets of areas as peripheral in the global scenario of cortical processing (notice also how the nodes in these two communities all have low values of HD (Table~\ref{table-hd})). In contrast, communities 1 and 3 seem to mediate information exchange and promote integration across the entire cortex. The higher position of their constituent areas in Table~\ref{table-hd} indicates that they do so by means of high-level information processing. Interestingly, this scenario is remarkably similar to the bow-tie structure suggested in \cite{markov13counterstream}.

\section{Reciprocal connections}\label{reciprocals}

Two connections are said to be reciprocal if they involve the same pair of areas but have opposite directions. The majority ($\sim 80\%$, totaling 214 pairs) of the interareal connections in the dataset are reciprocal \cite{markov12}, something that has been consistently observed in other datasets as well \cite{scannell95,modha10,oh14}. 

The idea of countercurrent streams of information, or of closed feedback loops, has long seemed only natural in a system that shows a great capacity of self-regulation \cite{ashby62}. This can happen by means of directed cycles in the cortical network, and the smallest kind of cycle possible is the one formed by reciprocal connections, which allows for an extremely fast feedback response (if one considers the number of hops needed to complete the cycle). Hence, one would expect a supposedly highly self-regulatory and efficient system such as the mammalian cortex to exhibit as much reciprocal connectivity as possible. We examine some of the characteristics pertaining to this class of connections in the macaque cortex.

\subsection{NN and FLN}

\begin{figure}[p]
\centering
\scalebox{0.80}{\includegraphics{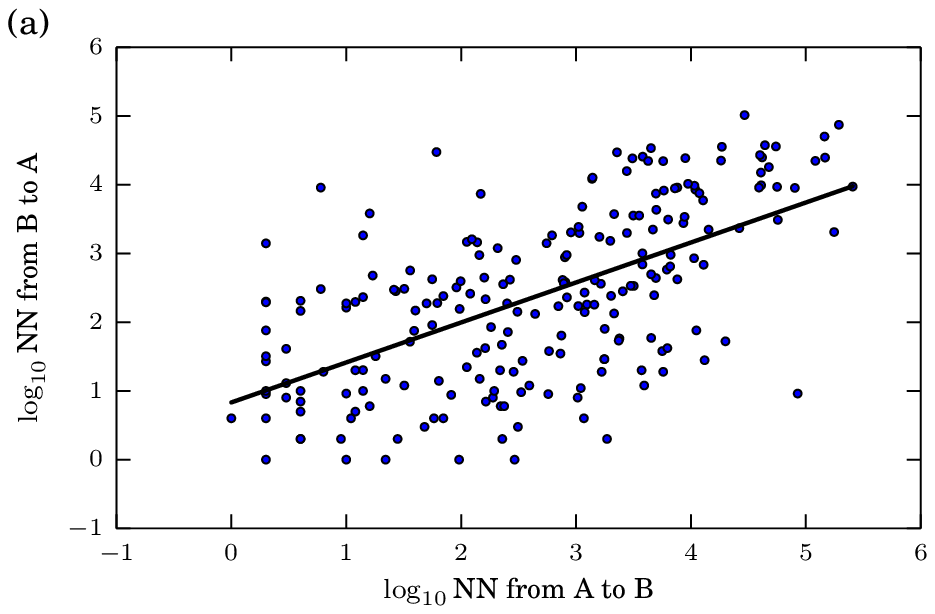}}
\scalebox{0.80}{\includegraphics{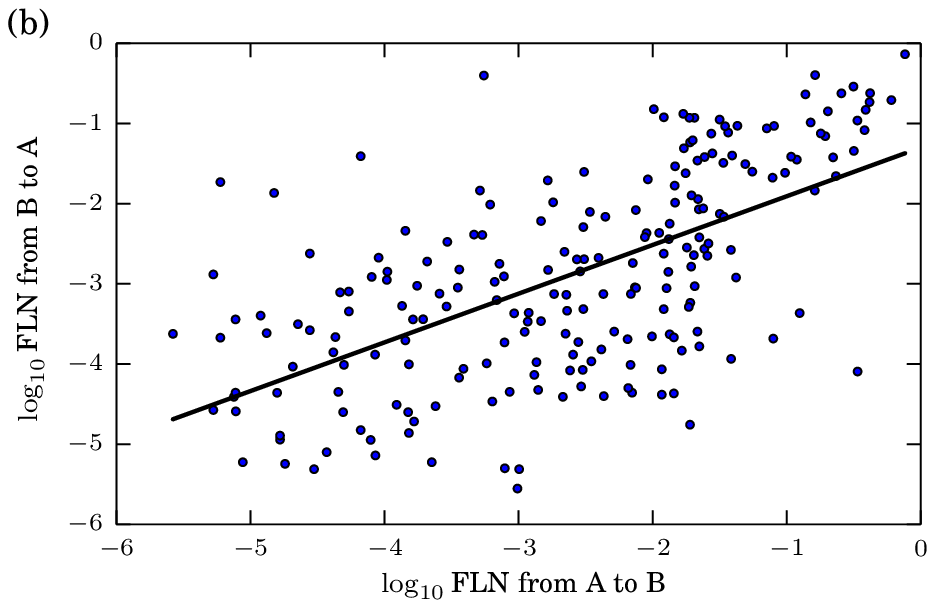}}
\scalebox{0.80}{\includegraphics{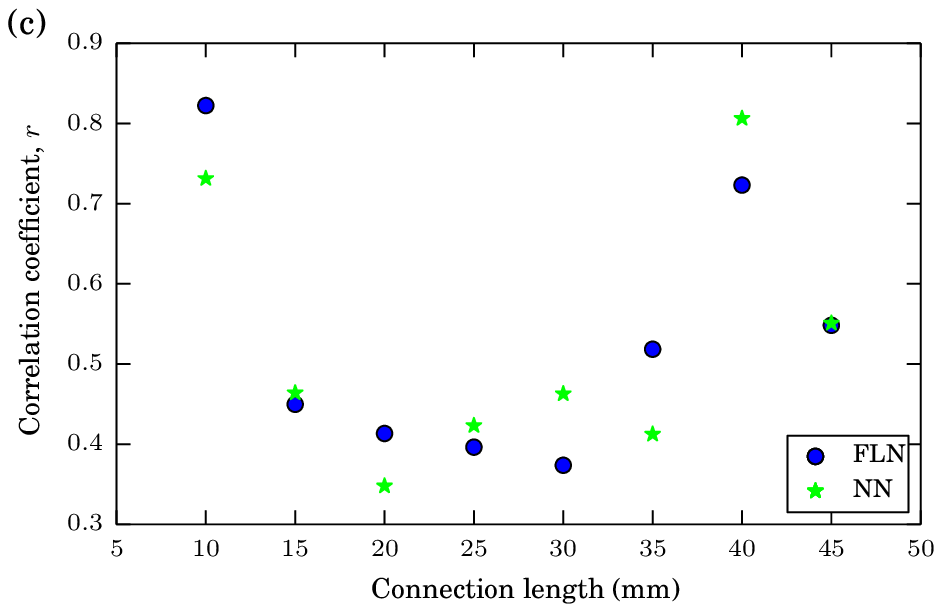}}
\caption{Correlation between the $\log_{10}\textrm{NN}$ of two reciprocal connections (a) and between their $\log_{10}\textrm{FLN}$ (b). In (a) and (b), each point corresponds to one of the 214 reciprocal-connection pair. Correlation coefficients ($r$) are shown for each of eight non-overlapping connection-length intervals (c), each represented by its rightmost value (a further interval, [45,50), comprises only two reciprocal-connection pairs and is for this reason omitted).}
\label{fig-recipr}
\end{figure}

We have found that reciprocal connections exhibit a strong correlation between their NN, as well as FLN, values in each direction. Figure~\ref{fig-recipr}(a) shows the result of plotting, for all 214 reciprocal-connection pairs, $\log_{10}\textrm{NN}$ in one direction \textit{vs.}\ $\log_{10}\textrm{NN}$ in the other direction. The same plot is found in Figure~\ref{fig-recipr}(b) but using FLN instead of NN. To decide which direction to use as abscissa in the plots, for each pair of areas we chose the direction from the lower to the higher area in the cortical hierarchy of Table~\ref{table-hd}. (For example: the pair (V1,V2) has reciprocal connections and V1 is lower in the hierarchy (smaller HD), therefore the pair's abscissa refers to $\textrm{V1}\rightarrow\textrm{V2}$ and its ordinate to $\textrm{V2}\rightarrow\textrm{V1}$.) Linear correlations were found for both NN logarithms ($r = 0.60$) and FLN logarithms ($r = 0.62$).

We also checked whether these correlations were valid over different ranges of connection length between the areas. Figure~\ref{fig-recipr}(c) shows that the correlation coefficients are high for short- as well as for long-distance reciprocal connections.

Note that the fact that NN values in reciprocal connections are correlated does not automatically imply that their FLN values should be correlated as well. Since the FLN of a given connection reflects the relative contribution of that connection to its target area, and since reciprocal connections have different targets, it would be perfectly possible for two reciprocal connections having similar NN values to contribute very differently to their target areas, and hence have very distinct FLN values.

Because of the general tendency of NN values to fall with increasing connection length between areas \cite{ercsey13}, one might suppose that, since reciprocal connections have approximately the same connection length, it is only expected that they should have correlated NN values. In fact, $\log_{10}\textrm{NN}$ does correlate negatively with connection length ($r = -0.48$, plot not shown). However, when we eliminate the effect of connection length by computing the partial correlation between $\log_{10}\textrm{NN}$ values in opposite directions, we still get a significant correlation ($r = 0.48$). This suggests that connection length, though playing a role in determining NN values, is not wholly determinant. The same reasoning is valid for FLN values as well. We return to this point in Section~\ref{discussion}.

\begin{figure}[t]
\centering
\scalebox{0.80}{\includegraphics{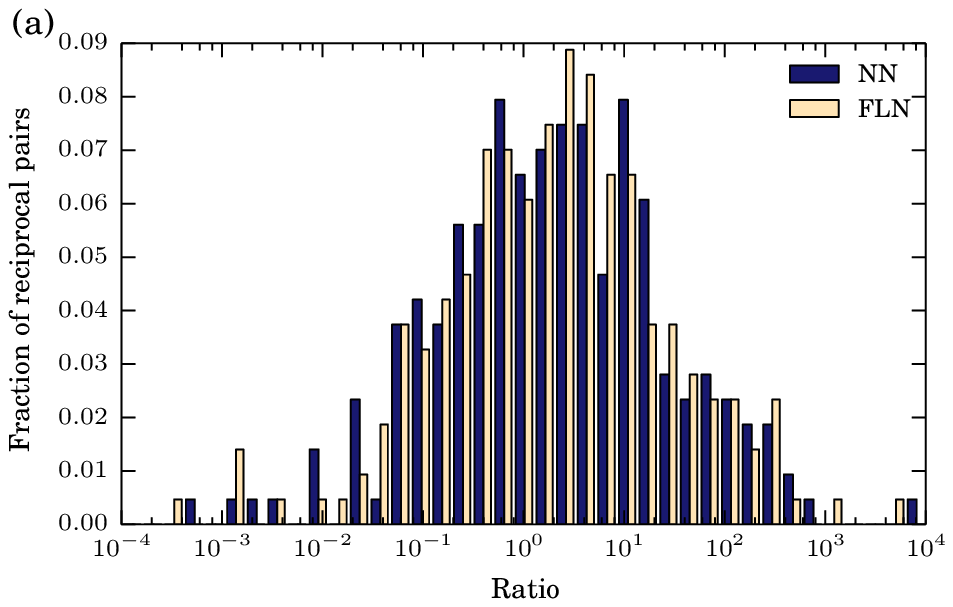}}
\scalebox{0.80}{\includegraphics{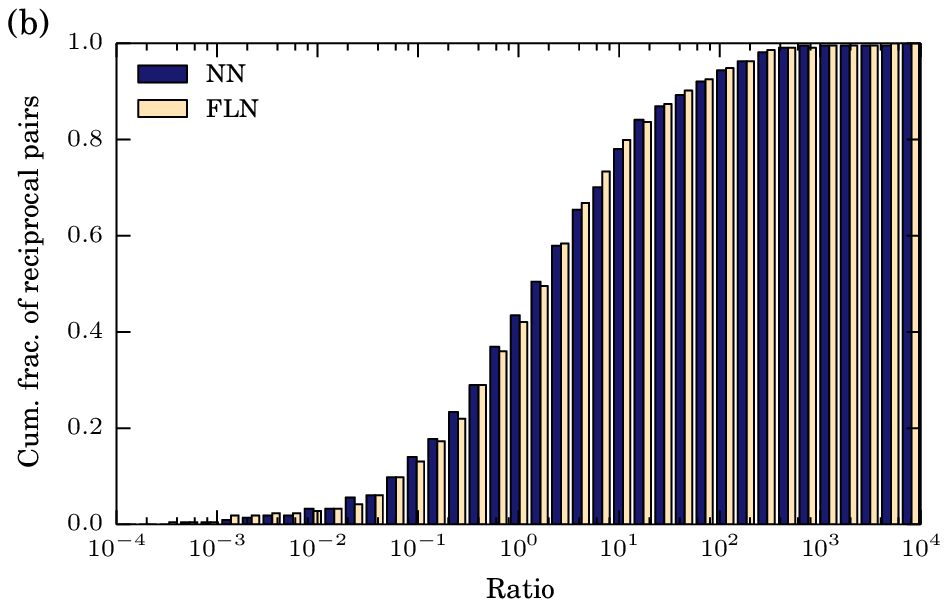}}
\caption{Distribution (a) and cumulative distribution (b) of NN and FLN ratios for reciprocal pairs of connections. Each ratio is given for the upward connection relative to the downward connection, where directions refer to the hierarchy of Table~\ref{table-hd}. Data are log-binned to the base 1.6.}
\label{fig-ratios}
\end{figure}

Not only do Figures~\ref{fig-recipr}(a),(b) show that there is a linear relation between the logarithms of opposite-direction NN or FLN values, but more importantly, the corresponding trend lines suggest that the values themselves might have similar orders of magnitude. To assess this, we have computed, for each pair of reciprocal connections, the ratio between the NN value in the direction of the lower to the higher area in the hierarchy of Table~\ref{table-hd} and the NN value in the opposite direction. The same was calculated for FLN values. Figures~\ref{fig-ratios}(a),(b) show the distribution and cumulative distribution, respectively, of the ratios of NN and FLN values for the 214 reciprocal pairs (the mean NN ratio is 69.3 and the mean FLN ratio is 45.6). Two important facts can be observed: the first is that the majority of reciprocal connections have NN or FLN values of not too different orders of magnitude in both directions, differing only by a factor of less than 100; the second is that the distribution of ratios is skewed to the right---connections from lower to higher areas in the cortical hierarchy usually employ more neurons than the connections in the opposite direction.

\subsection{Non-reciprocal connections}

\begin{figure}[t]
\centering
\scalebox{0.80}{\includegraphics{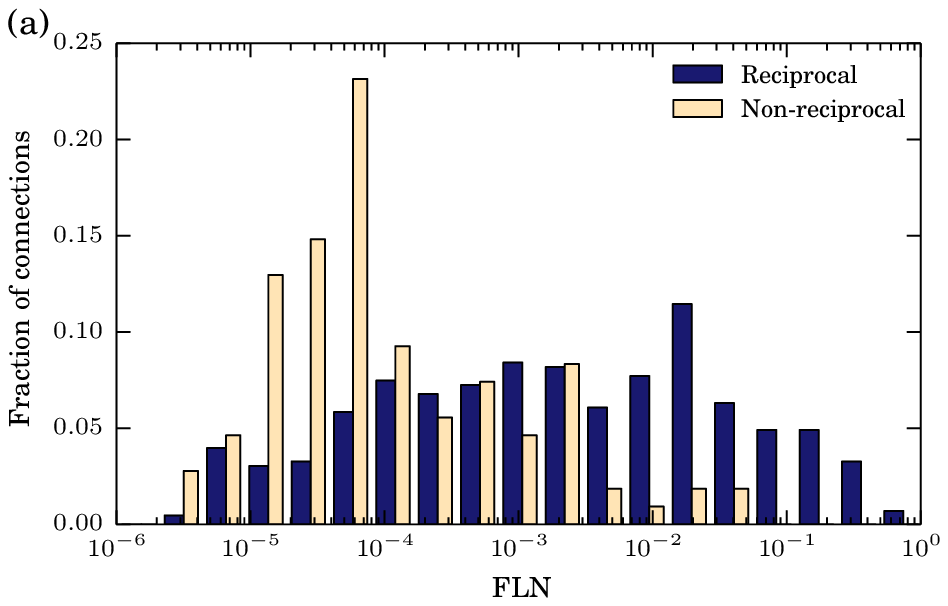}}
\scalebox{0.80}{\includegraphics{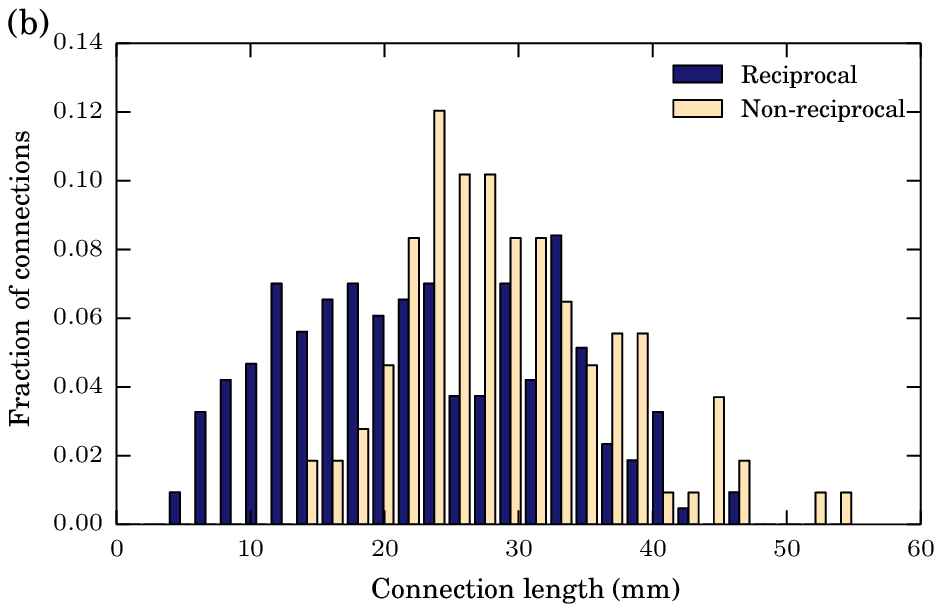}}
\caption{Comparison of reciprocal and non-reciprocal connections with respect to the distribution of FLN values (a) and of connection lengths (b). Data in panel (a) are log-binned to the base 2.}
\label{fig-nonrecipr}
\end{figure}

The results above elicit the question of what characterizes the connections that do not have a reciprocal counterpart, i.e., exist only from A to B but not from B to A. We have found them to comprise mainly connections with relatively small FLN values. This is illustrated in Figure~\ref{fig-nonrecipr}(a), which compares reciprocal and non-reciprocal connections with respect to their FLN distributions. This might at first suggest that maybe weaker connections are not very relevant, and thus dispense with the need for any reciprocity. However, a considerable member of the reciprocal connections are themselves weak (50\% of the reciprocal connections are among the 25\% weakest when considering the $29\times29$ adjacency matrix), which prevents us from making this generalization.

A second justification for non-reciprocity one might think of is that non-reciprocal connections are mostly long-distance, hence with high wiring costs due to connection length, making the absence of a reciprocal counterpart an energy-saving issue. But that is not strictly the case either, because, as can be seen in Figure~\ref{fig-nonrecipr}(b), there is a considerable portion of short-distance non-reciprocal connections.

Another possibility would be that non-reciprocal connections involve so few neurons that they could not all be detected by the experiments conducted, which also does not seem very likely given the consistency of the experimental results reported in \cite{markov12}.
Yet another justification for their existence is that perhaps they perform a fundamentally different function that does not require the kind of two-way signal exchange seen for the majority of connections. In fact, it has been suggested that some of them may be involved in direct top-down access to memory storage \cite{knierim92}, so maybe the investigation of this seemingly special kind of connection will benefit from further analysis that takes into full consideration the identities of the cortical areas involved.

\section{Discussion}\label{discussion}

Perhaps one of the most difficult aspects in the analysis of data from real networks is their interpretation. It is often difficult to assess the actual relevance from a biological point of view of many of the graph-theoretic properties typically investigated in such studies, such as the betweenness centrality of a given node, or the network's motif distribution, for example. Many of these properties have been thoroughly described for brain networks, and yet little has actually been revealed to provide us with insights into how information processing in the brain takes place that had not already been gained using other approaches.

In this study, we have proposed a hierarchical distance measure to be able to show that there seems to be a strong relation between the distribution of the weights of a given area's incoming connections and its position in the cortical processing hierarchy. The choice of the inverse of an edge's FLN as its length, though somewhat arbitrary, ensures that the more an area A is influenced by an area B (as reflected by the FLN value of link $\textrm{B}\rightarrow\textrm{A}$), the closer their position in a processing hierarchy should be.

We have also found the detection of link communities in cortical networks to be  an invaluable resource, exposing not only a main community structure analogous to a traditional node community but also the major flows of information between them. Many nodes, or areas, appear in many communities, indicating that each area, despite being a single entity with particular cytoarchitectural characteristics, probably has the capacity of acting differently and selectively according to the neighbor it is interacting with. From this point of view, as opposed to performing a single type of computation, which in the absence of other areas would be of no use (i.e., like a mechanical part in an automobile), a cortical area looks more like a processing unit on its own, with many levels of processing taking place depending on the sources of the incoming information, and, together with the other areas, makes possible an even greater integrated unit, namely the whole cortex. This local autonomy at the areal level provides for a global flexibility at the cortical level that might explain, for example, some of the remarkable feats of recovery of the nervous system after trauma or lesion.

The communities that were found present us with a general map of the major pathways of information in the macaque cortex. Even for such a small network in terms of number of nodes, its high link density makes direct inspection extremely difficult, even after filtering out the links with smallest weights. Link-community detection, therefore, has proven instrumental as a principled way of making sense of the intricate patterns found in such a dense network.

Finally, we have investigated the characteristics of reciprocal connections, finding them to be highly correlated with respect to NN logarithms as much as FLN logarithms in each direction. Moreover, the distribution of NN ratios for reciprocal-connection pairs tells us that reciprocal connections employ a similar number of neurons in each direction, and the distribution of FLN ratios tells us that reciprocal connections tend to have similar degrees of influence in each direction.

At this point is seems important to stress the difference between NN and FLN values. Even though the FLN values look like a natural way to translate projection data into edge weights, the NN values by themselves are also important quantities, since they approximately reflect the bandwidth of communication used to transmit signals between two areas. Therefore, FLN values tell only part of the story behind interareal communication in the cortex.

Having said this, one might inquire about the reasons why a given connection from A to B should involve a value of NN similar to that of the reciprocal connection from B to A. First note that, in principle, there is nothing to prevent some area A from using thousands of neurons to send signals (say, incoming visual inputs) to B, while receiving a response fired by just a few neurons in B, encoding something like a general-purpose reinforcing or modulatory message. After all, neural networks are extremely flexible in that they allow for all sorts of architectures, e.g., a single action potential fired by a single neuron can be relayed and made to propagate over a whole array of neurons.

Nevertheless, our results suggest that this is not what happens in the communication between two areas with reciprocal information channels: a response to signaling on a channel with a bandwidth of thousands of neurons will likewise employ thousands of neurons as well. This allows for a much finer response. To make a simplified analogy, it is the difference between, after reading a book, being able to give one's opinion about it by choosing between 1 to 5 stars or by writing an in-depth review, where one can specify which parts of the book were thought to be good or bad. With this kind of richer response, a cortical area can act upon different sections of the message received, inhibiting, modulating, or reinforcing it according to the organism's current needs and objectives. Therefore, the ubiquity of reciprocity in the communication between pairs of areas and the tendency toward relative symmetry help to underline the role of each cortical area as a largely self-contained unit, as discussed above.

We have also seen in Figure~\ref{fig-ratios} that connections from lower to higher areas in the cortical hierarchy tend to have slightly higher NN and FLN values than their reciprocal counterparts. This indicates that, even though the magnitudes of reciprocal connections are similar, less processed information needs a larger bandwidth than highly processed responses. One could imagine raw information being conveyed from lower to higher areas, increasingly being refined and more efficiently encoded along the way, eventually resulting in optimized feedback from the higher areas that nonetheless retains the essential characteristics of the original signal, as suggested by the similar opposite-direction magnitudes of NN and FLN values.

\subsection{Strength \textit{vs}.\  importance of connections}

In our analyses, we have filtered out the links with smallest FLN values to allow for better link-community detection, as well as for a more homogeneous ratio between NN values in reciprocal connections. We have also found a stronger correlation between mean FLN values and hierarchical distances when considering only the ten highest FLN values for each area.
This might lead us to question the importance of the weakest connections for cortical processing, since apparently they only obfuscate some useful analyses of the network, even though they correspond to more than 40\% of the connections present in the dataset (if we postulate the threshold of 0.000362 used above as a rough limit between strong and weak connections).

\begin{figure}[t]
\centering
\scalebox{0.80}{\includegraphics{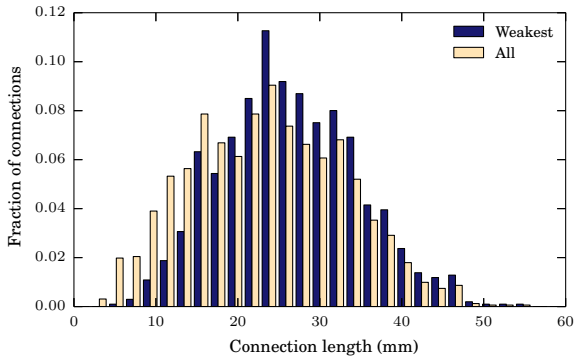}}
\caption{Distribution of connection lengths considering all connections and only the weakest ones ($\textrm{FLN} < 0.000362$).}
\label{fig-weakdists}
\end{figure}

Elsewhere these weaker connections have been associated with large connection lengths, suggesting an integrative role between cortical regions with distinct functions \cite{markov13counterstream} and the promotion of synchronous activity throughout the whole brain. It has also been suggested that they are responsible for connecting areas with distinct neighborhoods \cite{goulas14}. Nevertheless, weak connections exist over many connection-length ranges (Figure~\ref{fig-weakdists}), even inside communities where all areas are supposed to perform similar functions (as can be seen in the communities presented in Figure~\ref{fig-comms}).

One might thus be tempted to dismiss them as being secondary to the essential functioning of the brain. However, since we are still far from  understanding the exact role of each white-matter pathway in the communication between cortical regions, it would probably be erroneous to qualify a given connection as being less significant than another. Yet, if we think of neurons as a means for information transfer, we can interpret axonal projections as information channels, in which case it seems reasonable to expect a projection involving orders of magnitude more neurons than another to have a greater influence on the activity of the target area.

One might also advocate that the intrinsic importance of an interareal connection lies in the actual meaning of the signal being transmitted and its function toward the organism's survival. But even from this perspective, it seems hard to overlook the fact that, because of a simple matter of proportions, a connection involving 1000 times more neurons than another has the potential either to produce a greater physical effect or to have a larger repertoire of different effects and surely encode much more information. We also know that most of our biological apparatus is structured the way it is due to evolutionary pressures, so it should not be unreasonable to assume that connections employing more neurons have had a greater role in the organism's adaptation through the ages than a connection that is orders of magnitude ``thinner.''

On the other hand, given the considerable energy cost of communication in the brain, it seems unlikely that such a large number of connections would play no important part in cortical processing. Maybe weak connections have in fact a crucial role in cortical processing: they may facilitate synchronization and integration, or they may convey some specific kind of information, have a role in long-term memory recall, or even serve as alternative pathways to be used in case of lesion of malfunction in the brain. Future studies that focus on this specific aspect of cortical communication will definitely bring us one step further towards a better understanding of the mammalian brain.

\section{Conclusion}\label{conclusion}

In this study we have shed new light on some aspects of the interareal connectivity in the macaque cortex. First, we showed that the distribution of FLN values in a given area is indicative of that area's role in the cortical hierarchy, with mean FLN values being negatively correlated with hierarchical distances. Also, we have presented a way to assess the hierarchical distance of an area by computing the smallest weighted distance to the area in question from the primary sensory cortical areas. This distance can be used to compare the relative hierarchical positions of two given areas without the need of laminar distribution data.

Second, the detection of link communities confirmed a prominent modular organization in the macaque cortex. This technique proved particularly  adequate for revealing community structure in cortical networks, given that it naturally incorporates the overlapping of functions among areas. The directionality of the links in each community revealed the major flows of information in the network.

Third, there is a remarkable regularity in the NN and FLN values involved in reciprocal connections. Their ratios indicates that most reciprocal connections use communication bandwidths of the same order of magnitude and also have a similar relative importance for their target areas, with slightly higher values in the direction from lower to higher areas in the cortical hierarchy. In contrast, any justification of the existence of non-reciprocal connections seems to be still speculative.

These results were possible thanks to the specific characteristics of the quantitative tract tracing data made public in \cite{markov12}. Other connectome mapping techniques may be more practical but do not provide the same level of detail (or maybe express different aspects of cortical processing in the macaque brain).

Our findings will greatly benefit from more data, but at any rate they at least highlight pertinent aspects of cortico-cortical communication to be further investigated. We hope these results will encourage not only the production of more data but also an approach to brain network mapping and analysis that is more revealing of the issues still to be clarified. These include the precise role of weak and non-reciprocal connections and how the different cortical layers are used in interareal communication and intra-areal computations. We also speculate that, in the same way that the FLN distribution is related to cortical hierarchy, there may also be structural connectivity aspects enabling the comparison of areas on an evolutionary time scale.

\subsection*{Acknowledgments}

The authors acknowledge partial support from CNPq, CAPES, and a FAPERJ BBP grant.

\bibliography{paper}

\begin{thebibliography}{10}

\bibitem{ahn10}
Y.-Y. Ahn, J.~P. Bagrow, and S.~Lehmann.
\newblock Link communities reveal multiscale complexity in networks.
\newblock {\em Nature}, 466:761--764, 2010.

\bibitem{ashby62}
W.~R. Ashby.
\newblock Principles of the self-organizing system.
\newblock In H.~{von Foerster} and G.~W. {Zopf, Jr.}, editors, {\em Principles
  of Self-Organization: Transactions of the {U}niversity of {I}llinois
  Symposium}, pages 255--278. Pergamon Press, New York, NY, 1962.

\bibitem{barone00}
P.~Barone, A.~Batardiere, K.~Knoblauch, and H.~Kennedy.
\newblock Laminar distribution of neurons in extrastriate areas projecting to
  visual areas {V}1 and {V}4 correlates with the hierarchical rank and
  indicates the operation of a distance rule.
\newblock {\em The Journal of Neuroscience}, 20:3263--3281, 2000.

\bibitem{bassett08}
D.~S. Bassett, E.~Bullmore, B.~A. Verchinski, V.~S. Mattay, D.~R. Weinberger,
  and A.~Meyer-Lindenberg.
\newblock Hierarchical organization of human cortical networks in health and
  schizophrenia.
\newblock {\em The Journal of Neuroscience}, 28:9239--9248, 2008.

\bibitem{bressler10}
S.~L. Bressler and V.~Menon.
\newblock Large-scale brain networks in cognition: emerging methods and
  principles.
\newblock {\em Trends in Cognitive Sciences}, 14:277--290, 2010.

\bibitem{bullmore09}
E.~Bullmore and O.~Sporns.
\newblock Complex brain networks: graph theoretical analysis of structural and
  functional systems.
\newblock {\em Nature Reviews Neuroscience}, 10:186--198, 2009.

\bibitem{ercsey13}
M.~Ercsey-Ravasz, N.~T. Markov, C.~Lamy, D.~C. {Van Essen}, K.~Knoblauch,
  Z.~Toroczkai, and H.~Kennedy.
\newblock A predictive network model of cerebral cortical connectivity based on
  a distance rule.
\newblock {\em Neuron}, 80:184--197, 2013.

\bibitem{fve91}
D.~J. Felleman and D.~C. {Van Essen}.
\newblock Distributed hierarchical processing in the primate cerebral cortex.
\newblock {\em Cerebral Cortex}, 1:1--47, 1991.

\bibitem{goulas14}
A.~Goulas, A.~Schaefer, and D.~S. Margulies.
\newblock The strength of weak connections in the macaque cortico-cortical
  network.
\newblock {\em Brain Structure and Function}, 2014.
\newblock To appear.

\bibitem{hagmann08}
P.~Hagmann, L.~Cammoun, X.~Gigandet, R.~Meuli, C.~J. Honey, V.~J. Wedeen, and
  O.~Sporns.
\newblock Mapping the structural core of human cerebral cortex.
\newblock {\em PLo{S} Biology}, 6:e159, 2008.

\bibitem{hilgetag10}
C.~C. Hilgetag and S.~Grant.
\newblock Cytoarchitectural differences are a key determinant of laminar
  projection origins in the visual cortex.
\newblock {\em NeuroImage}, 51:1006--1017, 2010.

\bibitem{kaiser07}
M.~Kaiser, R.~Martin, P.~Andras, and M.~P. Young.
\newblock Simulation of robustness against lesions of cortical networks.
\newblock {\em European Journal of Neuroscience}, 25:3185--3192, 2007.

\bibitem{knierim92}
J.~J. Knierim and D.~C. {Van Essen}.
\newblock Visual cortex: cartography, connectivity, and concurrent processing.
\newblock {\em Current Opinion in Neurobiology}, 2:150--155, 1992.

\bibitem{markov13counterstream}
N.~T. Markov, M.~Ercsey-Ravasz, D.~C. {Van Essen}, K.~Knoblauch, Z.~Toroczkai,
  and H.~Kennedy.
\newblock Cortical high-density counterstream architectures.
\newblock {\em Science}, 342:1238406, 2013.

\bibitem{markov12}
N.~T. Markov, M.~M. Ercsey-Ravasz, A.~R. Ribeiro~Gomes, C.~Lamy, L.~Magrou,
  J.~Vezoli, P.~Misery, A.~Falchier, R.~Quilodran, M.~A. Gariel, J.~Sallet,
  R.~Gamanut, C.~Huissoud, S.~Clavagnier, P.~Giroud, D.~Sappey-Marinier,
  P.~Barone, C.~Dehay, Z.~Toroczkai, K.~Knoblauch, D.~C. {Van Essen}, and
  H.~Kennedy.
\newblock A weighted and directed interareal connectivity matrix for macaque
  cerebral cortex.
\newblock {\em Cerebral Cortex}, 24:17--36, 2014.

\bibitem{markov14hierarchy}
N.~T. Markov, J.~Vezoli, P.~Chameau, A.~Falchier, R.~Quilodran, C.~Huissoud,
  C.~Lamy, P.~Misery, P.~Giroud, S.~Ullman, P.~Barone, C.~Dehay, K.~Knoblauch,
  and H.~Kennedy.
\newblock Anatomy of hierarchy: feedforward and feedback pathways in macaque
  visual cortex.
\newblock {\em Journal of Comparative Neurology}, 522:225--259, 2014.

\bibitem{meunier10}
D.~Meunier, R.~Lambiotte, and E.~T. Bullmore.
\newblock Modular and hierarchically modular organization of brain networks.
\newblock {\em Frontiers in Neuroscience}, 4:200, 2010.

\bibitem{modha10}
D.~S. Modha and R.~Singh.
\newblock Network architecture of the long-distance pathways in the macaque
  brain.
\newblock {\em Proceedings of the National Academy of Sciences {U}{S}{A}},
  107:13485--13490, 2010.

\bibitem{newman04}
M.~E.~J. Newman.
\newblock Analysis of weighted networks.
\newblock {\em Physical Review E}, 70:056131, 2004.

\bibitem{oh14}
S.~W. Oh, J.~A. Harris, L.~Ng, B.~Winslow, N.~Cain, S.~Mihalas, Q.~Wang,
  C.~Lau, L.~Kuan, A.~M. Henry, M.~T. Mortrud, B.~Ouellette, T.~N. Nguyen,
  S.~A. Sorensen, C.~R. Slaughterbeck, W.~Wakeman, Y.~Li, D.~Feng, A.~Ho,
  E.~Nicholas, K.~E. Hirokawa, P.~Bohn, K.~M. Joines, H.~Peng, M.~J. Hawrylycz,
  J.~W. Phillips, J.~G. Hohmann, P.~Wohnoutka, C.~R. Gerfen, C.~Koch,
  A.~Bernard, C.~Dang, A.~R. Jones, and H.~Zeng.
\newblock A mesoscale connectome of the mouse brain.
\newblock {\em Nature}, 508:207--214, 2014.

\bibitem{reid09}
A.~T. Reid, A.~Krumnack, E.~Wanke, and R.~K{\"{o}}tter.
\newblock Optimization of cortical hierarchies with continuous scales and
  ranges.
\newblock {\em NeuroImage}, 47:611--617, 2009.

\bibitem{scannell95}
J.~W. Scannell, C.~Blakemore, and M.~P. Young.
\newblock Analysis of connectivity in the cat cerebral cortex.
\newblock {\em The Journal of Neuroscience}, 15:1463--1483, 1995.

\bibitem{scannell99}
J.~W. Scannell, G.~A. P.~C. Burns, C.~C. Hilgetag, M.~A. O'Neil, and M.~P.
  Young.
\newblock The connectional organization of the cortico-thalamic system of the
  cat.
\newblock {\em Cerebral Cortex}, 9:277--299, 1999.

\bibitem{sporns10}
O.~Sporns.
\newblock {\em Networks of the Brain}.
\newblock The MIT Press, Cambridge, MA, 2010.

\bibitem{sporns13segregation}
O.~Sporns.
\newblock Network attributes for segregation and integration in the human
  brain.
\newblock {\em Current Opinion in Neurobiology}, 23(2):162--171, 2013.

\bibitem{stam12}
C.~J. Stam and E.~C.~W. van Straaten.
\newblock The organization of physiological brain networks.
\newblock {\em Clinical Neurophysiology}, 123:1067--1087, 2012.

\bibitem{vandenheuvel13}
M.~P. van~den Heuvel and O.~Sporns.
\newblock An anatomical substrate for integration among functional networks in
  human cortex.
\newblock {\em The Journal of Neuroscience}, 33:14489--14500, 2013.

\bibitem{vandenheuvel13hubs}
M.~P. van~den Heuvel and O.~Sporns.
\newblock Network hubs in the human brain.
\newblock {\em Trends in Cognitive Sciences}, 17:683--696, 2013.

\bibitem{zamora10}
G.~Zamora-L\'{o}pez, C.~Zhou, and J.~Kurths.
\newblock Cortical hubs form a module for multisensory integration on top of
  the hierarchy of cortical networks.
\newblock {\em Frontiers in Neuroinformatics}, 4:1, 2010.

\end{thebibliography}
\bibliographystyle{plain}

\appendix
\section{Methods}\label{methods}

The formula for calculating the similarity, $S$, between two edges $e_{ik}$ and $e_{jk}$ in an weightless undirected network with no self-loops is
\begin{equation}
S(e_{ik},e_{jk}) = \frac{|n_+(i) \cap n_+(j)|}{|n_+(i) \cup n_+(j)|},
\label{firsteq}
\end{equation}
where $n_+(i)$ is the set of neighbors of node $i$, including $i$ itself (or, simply, the set of so-called inclusive neighbors of $i$). The reason for the inclusion of the node itself in its own neighborhood is clear when nodes $i$ and $j$ share the exact same neighbors: without the inclusion, $S$ would be equal to 1 irrespective of the actual number of common neighbors.

The presence of weights on the edges can be accounted for by using a vector form of Equation~(\ref{firsteq}),
\begin{equation}
S(e_{ik},e_{jk}) = \frac{a_i \cdot a_j}{|a_i|^2 + |a_j|^2 - a_i \cdot a_j},
\end{equation}
where $a_i$ is the $i$th row of the weighted adjacency matrix (except for the diagonal element). That is, if $N$ if the number of nodes in the network, then $a_i = (w_{i1}, w_{i2}, \dots, w_{iN})$, each element standing for the weight of the corresponding edge (except for $w_{ii}$, which we discuss next).

To be coherent with the use of inclusive neighbors in Equation~(\ref{firsteq}), we need to decide upon an adequate value for $w_{ii}$. In \cite{ahn10} it is proposed that this element be the arithmetic mean of the weights of the edges incident to $i$. We make use of an illustrative example to suggest that a better option is the maximum weight among the edges incident to $i$.

Consider a simple network composed of three nodes, $i$, $j$, and $k$, and of two edges, $e_{ik}$ and $e_{jk}$, both with unit weights. Since $i$ and $j$ have no neighbor in common other than $k$, we have $S(e_{ik},e_{jk}) = 0.333$, as $w_{ii} = 1$ and $w_{jj} = 1$ no matter which method we use for handling self-weights. However, if we add a node $l$ and an edge $e_{il}$ to this network, the two methods yield different values of $S$($e_{ik}$, $e_{jk}$), depending on the weight $w_{il}$. If $w_{il} = 1$, both methods yield $S(e_{ik},e_{jk}) = 0.25$---which makes perfect sense, since $i$ now has a neighbor not adjacent to $j$. But if we let $w_{il} = 0.1$, for example, using the mean weight $w_{ii} = (1+0.1)/2 = 0.55$ results in $S(e_{ik},e_{jk}) = 0.43$, a value even greater than the original similarity of 0.333 when $i$ and $j$ had the same non-inclusive neighborhood. Using the maximum weight, $w_{ii} = \max\{1,0.1\} = 1$, results in  $S(e_{ik},e_{jk}) = 0.332$, a slightly smaller value due to the new edge introduced. In other words, the use of the maximum weight reflects asymmetries in the neighborhoods of two nodes in a way that better takes into account the magnitude of the weights in those neighborhoods. Therefore, we henceforth use
\begin{equation}
w_{ii} = \max_{i' \in n(i)} w_{ii'},
\end{equation}
where $n(i)$ is the (non-inclusive) neighborhood of $i$.

Now, to incorporate directions into the above calculations, we believe that neighbors with the same identity but interacting in different directions should be treated as distinct neighbors altogether. For imagine we are calculating the similarity between two directed edges $e_{ik}$ and $e_{jk}$, and that both $i$ and $j$ have $l$ as a neighbor, but in different directions (e.g., $l$ is an in-neighbor of $i$ and an out-neighbor of $j$). The similarity value should be smaller than if $l$ interacted with $i$ and $j$ in the same direction.

Therefore, a straightforward way to adapt Equation~(\ref{firsteq}) to the presence of directions is to make
\begin{equation}
S(e_{ik},e_{jk}) = S(e_{ki},e_{kj}) = \frac{|n_+^\mathrm{in}(i) \cap n_+^\mathrm{in}(j)| + |n_+^\mathrm{out}(i) \cap n_+^\mathrm{out}(j)|}
{|n_+^\mathrm{in}(i) \cup n_+^\mathrm{in}(j)| + |n_+^\mathrm{out}(i) \cup n_+^\mathrm{out}(j)|},
\end{equation}
or, in vector form,
\begin{equation}
S(e_{ik},e_{jk}) = \frac{a_i^\mathrm{in} \cdot a_j^\mathrm{in} + a_i^\mathrm{out} \cdot a_j^\mathrm{out}}
{|a_i^\mathrm{in}|^2 + |a_j^\mathrm{in}|^2 - a_i^\mathrm{in} \cdot a_j^\mathrm{in} + |a_i^\mathrm{out}|^2 + |a_j^\mathrm{out}|^2 - a_i^\mathrm{out} \cdot a_j^\mathrm{out}},
\end{equation}
where both edges have the same direction (i.e., toward $k$ or away from $k$).

As for edge pairs having different directions with respect to the node they have in common, none of them is taken into account in our calculations. We made this decision because it seems to us that such edges should not have similarities greater than any pair of edges incident to a common node in the same direction.

Once we calculated the similarity values for the relevant edge pairs in the dataset, we proceeded by applying single-linkage hierarchical clustering to construct a link dendrogram (with ties in similarity value being incorporated simultaneously). The resulting communities were selected at the point of maximum partition density, $D$, as proposed in \cite{ahn10}. This density is given by
\begin{equation}
D = \frac{1}{M}\sum_c m_c D_c,
\end{equation}
where $M$ is the network's number of edges, $c$ ranges over all communities, and $D_c$, following a straightforward adaptation to the directed case, is given by
\begin{equation}
D_c = \frac{m_c - (n_c - 1)}{n_c(n_c - 1) - (n_c - 1)}.
\end{equation}
In this expression, $m_c$ and $n_c$ are, respectively, the number of edges and nodes in community $c$. So $D$ is the average $D_c$ value, each community weighted by the fraction of $M$ to which its edges correspond. To understand the meaning of the quantity $D_c$, note that $n_c - 1$ is the minimum number of edges required for $n_c$ nodes to be connected. So $D_c$ can be seen as the number of edges community $c$ has in excess of this minimum, normalized to the maximum excess there can be (i.e., when all possible $n_c(n_c - 1)$ directed edges are present). A detailed discussion of link communities is given in \cite{ahn10}.

\end{document}